# When learning analytics dashboard is explainable: An exploratory study on the effect of GenAI-supported learning analytics dashboard

Angxuan Chen
Department of Educational Technology
Peking University
Beijing China
Angxuan.chen@stu.pku.edu.cn

## ABSTRACT

This study investigated the impact of a theory-driven, explainable Learning Analytics Dashboard (LAD) on university students' human-AI collaborative academic abstract writing task. Grounded in Self-Regulated Learning (SRL) theory and incorporating Explainable AI (XAI) principles, our LAD featured a three-layered design (Visual, Explainable, Interactive). In an experimental study, participants were randomly assigned to either an experimental group (using the full explainable LAD) or a control group (using a visual-only LAD) to collaboratively write an academic abstract with a Generative AI. While quantitative analysis revealed no significant difference in the quality of co-authored abstracts between the two groups, a significant and noteworthy difference emerged in conceptual understanding: students in the explainable LAD group demonstrated a superior grasp of abstract writing principles, as evidenced by their higher scores on a knowledge test (p= .026). These findings highlight that while basic AI-generated feedback may suffice for immediate task completion, the provision of explainable feedback is crucial for fostering deeper learning, enhancing conceptual understanding, and developing transferable skills fundamental to self-regulated learning in academic writing contexts.

## 1 Introduction

Learning Analytics Dashboards (LADs) have been recognized as important instructional tools and have attracted considerable research attention over the past decade [1]. Numerous studies suggest that LADs can be effective in certain contexts, for instance, by providing students with feedback on their learning processes, promoting self-reflection, and potentially improving learning outcomes[5]. They are often designed to help students monitor their progress and engagement, which can support Self-Regulated Learning (SRL) [7].

However, existing research also highlights a significant limitation: LADs do not always adequately support student learning [2]. A primary reason for this is that the information presented on dashboards may not always be understood, accepted, or acted upon by students. Students might find the data unclear, irrelevant to their needs, or they may not trust the insights provided [8]. This lack of student buy-in or comprehension can occur if the dashboard functions as a "black box," where students do not understand how the data is collected, analyzed, or why specific recommendations are made (e.g., Sedrakyan et al., 2020). When students cannot connect the dashboard's information to their own learning experiences and goals, the potential of LADs to empower them is reduced.

To address these challenges, our research focuses on designing and developing a theory-driven and explainable Learning Analytics Dashboard. The design is 'theory-driven' because it is grounded in Self-Regulated Learning theory [9]. This ensures that the feedback and support offered are pedagogically sound and aligned with how students learn. More importantly, our dashboard incorporates 'explainability,' drawing from principles of Explainable AI (XAI) [4]. This means we aim to clearly communicate to students the logic behind the data indicators, the meaning of visualizations, and the basis for any personalized advice.

Therefore, by creating an LAD that is both theoretically grounded and transparent in its operations, this study seeks to advance research in Learning Analytics Dashboards. We aim to explore how integrating theoretical foundations with explainability can enhance student trust, comprehension, and ultimately, the effective use of LADs for learning. This work opens up new avenues for investigating how to make LADs more effective and truly student-centered tools.

## 2 Dashboard design

The Learning Analytics Dashboard (LAD) presented in this study is specifically designed to support students in human-AI collaborative academic writing tasks. The primary scenario involves students interacting and collaborating with a Generative AI (GenAI) to co-writing an academic abstract. Grounded in Self-Regulated Learning (SRL) theory [9], our dashboard is structured into several distinct modules (See Figure 1):

Abstract Writing Preparation Module: This module focuses on visualizing the learner's self-monitoring before they begin writing the abstract, aligning with the forethought and planning phase of Zimmerman's SRL model. Specifically, learners are prompted to set their target scores for five dimensions: Expected Time (min), Logical Coherence, Expression Accuracy, Structure Completeness, and Content Understanding. The selection of these evaluation dimensions is informed by the academic abstract writing rubric from previous study [6]. To assist learners in accurately self-monitoring their current writing proficiency and setting realistic goals, the dashboard provides reference examples for target scores within each dimension. This is visualized as a radar chart where students set their initial goals.

Writing Progress Monitoring Module: This module aims to visualize the completion status of various sub-sections of the abstract during the writing process, thereby directing students' attention to their ongoing performance. This corresponds to the performance phase of SRL. We have divided the abstract writing process into five key dimensions: Research Background, Research Question, Research Method, Research Results, and Research Conclusion. The dashboard provides a real-time visualization of progress in each of these areas using horizontal bar charts. The assessment of progress for each dimension is automated. Based on research indicating that well-prompted Large Language Models (LLMs) can achieve human-level accuracy in text evaluation and classification [3], a backend Deepseek-R1 model evaluates the completeness. The prompts for this model include the specific criteria for assessing progress in each dimension.

Writing Reflection Module: This module focuses on evaluating the student's completed abstract. It provides scores for Logical Coherence, Expression Accuracy, Structure Completeness, and Content Understanding, and also allows for comparison of actual time spent against the Expected Time. These scores are also generated by the backend DeepSeek-R1 model, using the previous academic abstract writing rubric as its evaluation standard [6]. The results are displayed on a radar chart. Crucially, this module overlays the student's initial target scores (from the Abstract Writing Preparation Module) onto the same radar chart. This direct visual comparison allows learners to intuitively see the gap between their self-set goals and their actual performance in each dimension, encouraging them to reflect on their work and identify areas for revision.

AI Dialogue Quality Monitoring Module: To support effective human-AI collaboration, this module visualizes the quality of the student's interaction with the GenAI. It analyzes the student's five most recent questions or prompts directed to the GenAI. The DeepSeek-R1 model assesses these interactions across four dimensions: (1) Task Focus: Whether the questions are focused on the abstract writing task. (2) Academic Standards: Whether the inquiries relate to academic writing conventions and quality. (3) Independent Thinking: Whether the student is seeking guidance and suggestions versus asking the AI to directly generate content. (4) Questioning Strategies: Whether the student employs effective strategies like asking follow-up questions or if there's a logical connection between consecutive prompts. The scores for these dimensions are presented as a bar chart.

To realize the 'explainability' aspect of our LAD, the interface is structured into three hierarchical layers, moving from visual overview to detailed, interactive explanations:

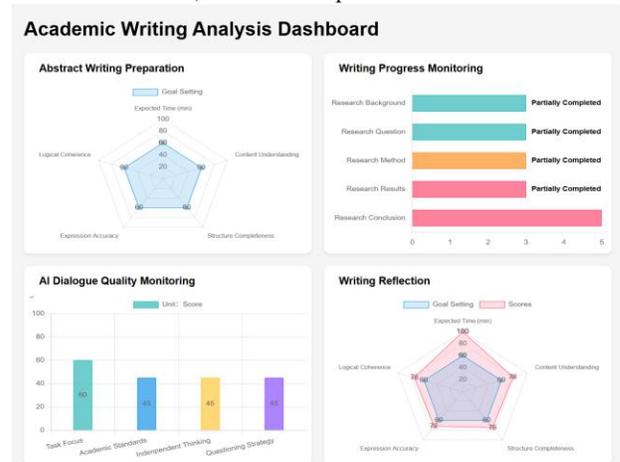

**Figure 1 Academic Writing Analysis Dashboard (Visual Layer)**

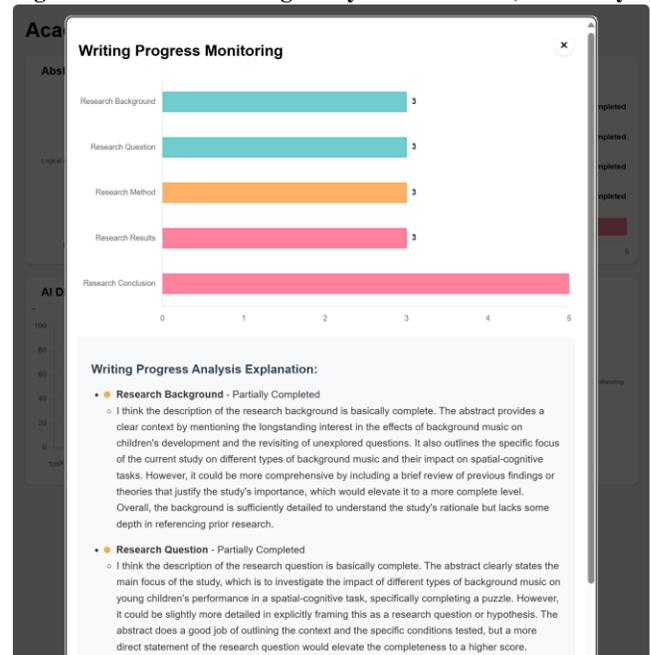

**Figure 2 Academic Writing Analysis Dashboard (Explainable Layer)**

1. Visual Layer (See Figure 1): This is the primary interface that displays all the aggregated and processed information through the visualizations described in the modules above (e.g., radar charts, bar graphs, progress bars). This layer provides an at-a-glance overview of performance and progress.

2. Explainable Layer (See Figure 2): This layer provides transparency into the AI's assessment process. When a student interacts with a specific score or metric (for instance, by clicking on a progress bar in the 'Writing Progress Monitoring' module to see "Writing Progress Analysis Explanation"), this layer reveals the chain of thinking (CoT) from the Deepseek-R1 model. It clearly shows why their abstract (or a component of it) received a particular score for a specific dimension. Furthermore, it offers actionable suggestions for improvement. This helps students understand the feedback and how to enhance their writing.

3. Interactive Layer (See Figure 3): This layer enables further dialogue and clarification. If students find the explanations provided in the Explainable Layer to be insufficient, unclear, or if they disagree with an assessment, they can select the relevant text within the explanation and pose follow-up questions. The backend Deepseek-R1 model is then used again to provide a more detailed and targeted response, further clarifying the basis for the score or suggestion. This interactive inquiry process aims to promote a deeper understanding of the feedback and foster student trust in the dashboard's information.

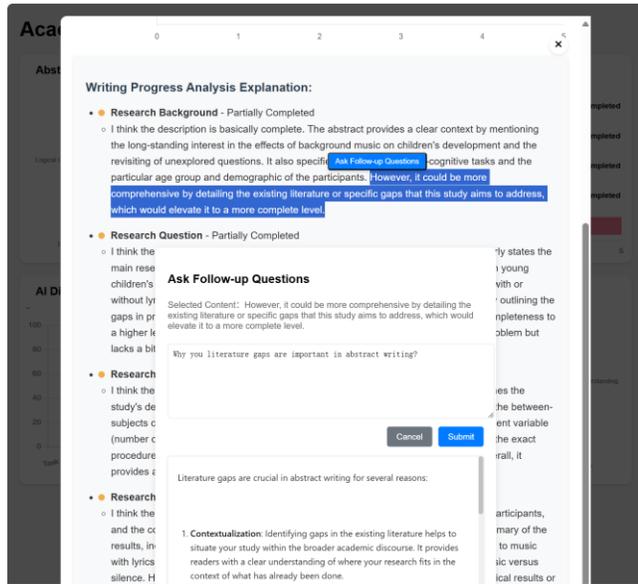

**Figure 3 Academic Writing Analysis Dashboard (Interactive Layer)**

## 3 Research Design

To investigate the impact of dashboard explainability on learners' performance in human-AI collaborative writing, an experimental study was conducted. We recruited thirty-nine non-native English-speaking university students who were then randomly assigned to one of two conditions: an experimental group (N=22) or a control group (N=17). Learners in the experimental group utilized the LAD equipped with all three layers: Visual, Explainable, and Interactive. In contrast, learners in the control group used an LAD that featured only the Visual layer. The core task for all participants was to write an academic abstract. They were provided with a research paper on music education, with its original abstract removed, to serve as the source material. Participants were instructed to collaborate with a Generative AI (GenAI) to compose the abstract for this paper. To facilitate this collaboration, a ChatAI module was integrated, powered by the DeepSeek-R1 model. The abstracts produced by the participants during the writing process were collected for subsequent analysis.

The written abstracts were quantitatively assessed for quality using a comprehensive scoring rubric. Each abstract was evaluated across seven criteria, with scores ranging from 0 to 3 points for each, based on the quality and presence of specific elements as detailed (See Figure 4):

| Abstract Criteria | 0 points | 1 point | 2 points | 3 points |
|---|---|---|---|---|
| Introductory Statement | Missing | Unclear; Doesn't connect to literature | Clear, but not engaging; Attempts to connect to literature | Clear, concise, engaging; Describes the topic to literature and purpose of work |
| Purpose | Missing | Unclear; Contains irrelevant or unimportant information | Clear but not concise; Might contain irrelevant or unimportant information; lacks specifics | Clear, concise, and relevant |
| Methodological Approach | Missing | Not mentioned but implied; or, not appropriate for purpose of scholarship | Unclear or not connected to purpose of scholarship | Connected to the purpose of the scholarship; Identifies method used to support thesis or answer the research question |
| Findings | Missing | Unclear; Or not related to the purpose of the scholarship; Or misinterpretation of results | Attempts to present findings but might be unclear; or some information missing | Clear, connected to the purpose of scholarship; Provides explanation of what was expected, discovered, accomplished, collected, produced |
| Contribution to Discipline | Missing | Unclear and lacks detail of contribution to the discipline | Attempts to connect work to discipline, but might be unclear | Clearly states how work advances knowledge in the discipline, why it's important, or how it can be used |
| Professional Writing | Grammatical errors, typos impede understanding; inappropriate verb tense | Many grammatical errors, typos but they do not impede understanding; inappropriate verb tense | Few grammatical errors or typos; Mixed verb tense | Writing appropriate for the profession; Defines all acronyms at first use; Appropriate verb tense (present/past tense when talking about the study, may use future tense for the contribution to the discipline. |
| Length | Too long or too short | | | 250-300 words |

**Figure 4 Abstract scoring rubric**

Additionally, after completing the collaborative writing task, participants undertook a knowledge test focused on abstract writing. This test comprised 10 multiple-choice questions (e.g., "Which of the following is not a main function of an abstract?") designed to assess their depth of understanding regarding the principles and conventions of abstract writing.

## 4 Result

### 4.1 Abstract Writing Performance

To compare the quality of abstracts written by learners in the experimental group (EG, N=22) and the control group (CG, N=17), independent samples t-tests were conducted. The analysis focused on seven specific dimensions of abstract quality as well as the total score.

The results indicated no statistically significant differences between the experimental group and the control group on any of the assessed dimensions of abstract writing. Specifically, there were no significant differences in scores for the Introductory Statement (t(37) = -1.840, p= .074), Purpose (t(37) = 0.856, p= .398), Methodological Approach (t(37) = -1.122, p= .269), Findings (t(37) = -0.062, p= .951), Contribution to Discipline (t(37) = -0.797, p= .431), Professional Writing (t(37) = 0.052, p= .959), or Length (t(37) = 0.140, p= .890).

Furthermore, the analysis of the total abstract scores (Sum Score) also revealed no significant difference between the two groups (t(37) = -0.650, p = .520). The mean total score for the control group was 14.94 (SD = 4.26), while the experimental group had a mean total score of 15.73 (SD = 3.30). These findings suggest that the type of dashboard used (either with full three-layer explainability or visual-only) did not differentially impact the overall quality of the abstracts produced by the learners. Both dashboard conditions appeared to similarly support learners in completing the abstract writing task.

### 4.2 Abstract Writing Knowledge Test

To compare understanding of abstract writing principles, scores on the abstract writing knowledge test were analyzed. Descriptive statistics showed that the experimental group (M = 90.00, SD = 6.90, N=22) had a higher mean score than the control group (M = 81.76, SD = 14.72, N=17). Given the violation of the homogeneity of variances assumption for a t-test (Levene's Test: $F(1, 37) = 6.622$, $p = .014$), a non-parametric Mann-Whitney U test was employed to compare the distributions of scores between the two groups.

The Mann-Whitney U test revealed a statistically significant difference in knowledge test scores between the experimental group and the control group ($W = 112.5$, $p = .026$). This result indicates that learners who used the dashboard with enhanced explainability features (experimental group) demonstrated a significantly better understanding of abstract writing principles compared to those who used the visual-only dashboard (control group). This suggests that the explainable dashboard may have contributed more effectively to a deeper conceptual understanding of abstract writing.

## 5 Discussion

This study investigated the impact of a theory-driven, explainable LAD on non-native English-speaking university students' academic abstract writing performance in human-AI collaboration and their understanding of abstract writing principles. The findings offer a nuanced perspective on the role of explainability in LADs.

A key finding was the absence of a significant difference in abstract writing quality between the experimental group (fully explainable LAD) and the control group (visual-only LAD). This suggests that for a single, relatively short writing task, even visual AI feedback provides enough guidance to help students produce comparable quality abstracts. Both dashboards, built on SRL theory, likely offered sufficient scaffolding for immediate task completion. It's plausible that improving tangible writing output, especially in complex academic genres, might require more prolonged engagement with the LAD or manifest across multiple writing tasks.

In contrast, a significant and noteworthy difference emerged in conceptual understanding: students interacting with the explainable LAD demonstrated a superior grasp of abstract writing principles, as evidenced by their higher scores on the knowledge test. This finding strongly supports our core premise: making AI-driven feedback transparent can foster deeper learning and comprehension. The explainable layer, by revealing the AI's "chain of thinking" and offering actionable suggestions, likely moved students beyond passive reception of scores. Instead, they could actively engage with the rationale behind the feedback, understand the criteria for effective abstract writing more profoundly, and critically reflect on their own work. The interactive layer further facilitated this by empowering them to seek clarification, promoting a more active and engaged learning process. This deeper engagement with the feedback process itself, facilitated by explainability, is the most probable reason for their enhanced conceptual understanding.

This distinction between task performance and knowledge acquisition is critical for understanding the broader educational potential of LADs. While immediate task improvement is a common goal, fostering transferable knowledge and skills is arguably more valuable. The explainable LAD appears to have achieved this more effectively. It didn't merely indicate areas for improvement; it helped students understand why and how to improve, which is fundamental for developing self-regulated learning skills [9]. Students equipped with this deeper understanding are better prepared to apply their knowledge to future, independent writing tasks, even without direct dashboard support.

In summary, the three-layered design of our LAD (Visual, Explainable, Interactive) offers a promising pedagogical framework. The visual layer provides an immediate overview, while the subsequent layers allow students to progressively deepen their understanding based on individual needs. This layered approach effectively manages cognitive load while still providing pathways for profound learning, aligning well with principles of effective instructional design.

## REFERENCES

[1] Naif Radi Aljohani, Ali Daud, Rabeeh Ayaz Abbasi, Jalal S Alowibdi, Mohammad Basheri, and Muhammad Ahtisham Aslam. 2019. An integrated framework for course adapted student learning analytics dashboard. *Computers in Human Behavior* 92, (2019), 679–690.

[2] Robert Bodily and Katrien Verbert. 2017. Review of research on student-facing learning analytics dashboards and educational recommender systems. *IEEE Transactions on Learning Technologies* 10, 4 (2017), 405–418.

[3] Dennis Foung, Linda Lin, and Julia Chen. 2024. Reinventing assessments with ChatGPT and other online tools: Opportunities for GenAI-empowered assessment practices. *Computers and Education: Artificial Intelligence* 6, (2024), 100250.

[4] Hassan Khosravi, Simon Buckingham Shum, Guanliang Chen, Cristina Conati, Yi-Shan Tsai, Judy Kay, Simon Knight, Roberto Martinez-Maldonado, Shazia Sadiq, and Dragan Gašević. 2022. Explainable artificial intelligence in education. *Computers and Education: Artificial Intelligence* 3, (2022), 100074.

[5] Wannisa Matcha, Nora'ayu Ahmad Uzir, and Dragan Ga. 2020. A Systematic Review of Empirical Studies on Learning Analytics Dashboards: A Self-Regulated Learning Perspective. *IEEE TRANSACTIONS ON LEARNING TECHNOLOGIES* 13, 2 (2020), 226–245.

[6] Edmond Sanganyado. 2019. How to write an honest but effective abstract for scientific papers. *Scientific African* 6, (2019), e00170.

[7] Gayane Sedrakyan, Jonna Malmberg, Katrien Verbert, Sanna Järvelä, and Paul A Kirschner. 2020. Linking learning behavior analytics and learning science concepts: Designing a learning analytics dashboard for feedback to support learning regulation. *Computers in Human Behavior* 107, (2020), 105512.

[8] Lixiang Yan, Linxuan Zhao, Vanessa Echeverria, Yueqiao Jin, Riordan Alfredo, Xinyu Li, Dragan Gaševi'c, and Roberto Martinez-Maldonado. 2024. VizChat: enhancing learning analytics dashboards with contextualised explanations using multimodal generative AI chatbots. In *International Conference on Artificial Intelligence in Education*, 2024. Springer, 180–193.

[9] B J Zimmerman. 2001. Theories of self-regulated learning and academic achievement: An overview and analysis. In *Self-Regulated Learning and Academic Achievement*. Lawrence Erlbaum Associates, Mahwah, NJ, USA, 1–37.